\documentstyle[aps,epsf,twoside,floats]{revtex}

\begin{document}

\markboth{\rm A. W\'ojs, K.-S. Yi, and J. J. Quinn}
         {Energy spectra and photoluminescence of FQH systems}
\pagestyle{myheadings}

\title{Energy spectra and photoluminescence of fractional 
       quantum Hall systems containing a valence-band hole}

\author{
   {\sc Arkadiusz W\'ojs}$\dagger\ddagger$,
   {\sc Kyung-Soo Yi}$\dagger\S$, and
   {\sc John J. Quinn}$\dagger$}

\address{\mbox{}\\[-2ex]\rm
   $\dagger$University of Tennessee, Knoxville, Tennessee 37996, USA \\
   $\ddagger$Wroclaw University of Technology, Wroclaw 50-370, Poland \\
   $\S$Pusan National University, Pusan 609-735, Korea}

\maketitle

\begin{abstract}
   The energy spectrum of a two-dimensional electron gas (2DEG) 
   interacting with a valence-band hole is studied in the high 
   magnetic field limit as a function of the filling factor $\nu$ 
   and the separation $d$ between the electron and hole layers.
   For $d$ smaller than the magnetic length $\lambda$, the hole 
   binds one or more electrons to form neutral ($X$) or charged 
   ($X^-$) excitons.
   The low-lying states can be understood in terms of 
   Laughlin-like correlations among the constituent charged 
   fermions (electrons and $X^-$).
   For $d$ comparable to $\lambda$, the electron--hole interaction 
   is not strong enough to bind a full electron, and fractionally 
   charged excitons $h$QE$_n$ (bound states of a hole and one or 
   more Laughlin quasielectrons, QE) are formed.
   The effect of these excitonic complexes on the photoluminescence 
   spectrum is studied numerically for a wide range of values 
   of $\nu$ and $d$.
\end{abstract}

\section{Introduction}
There have been many experimental studies of photoluminescence 
(PL) in fractional quantum Hall systems during the past decade 
(Heiman {\sl et al.}\ 1988,
Turberfield {\sl et al.}\ 1990,
Goldberg {\sl et al.}\ 1990,
Buhmann {\sl et al.}\ 1990, 1992, 1995, 
Goldys {\sl et al.}\ 1992, 
Kukushkin {\sl et al.}\ 1994,
Takeyama {\sl et al.}\ 1998,
Gravier {\sl et al.}\ 1998, 
Kheng {\sl et al.}\ 1993,
Shields {\sl et al.}\ 1995,
Finkelstein {\sl et al.}\ 1995, 1996, 
Hayne {\sl et al.}\ 1999,
Nickel {\sl et al.}\ 1998,
Tischler {\sl et al.}\ 1999,
Wojtowicz  {\sl et al.}\ 1999, 
Kim {\sl et al.}\ 2000,
Munteanu {\sl et al.}\ 2000,
Jiang {\sl et al.}\ 1998,
Brown {\sl et al.}\ 1996),
but the data have been rather difficult to interpret.
In order to obtain a more complete understanding of the PL process, 
it is essential to understand the nature of the low-energy states 
of the electron--hole system, and to evaluate their oscillator 
strength for radiative recombination.
In this note we investigate the elementary excitations of a system 
consisting of $N$ electrons ($e$) confined to the plane $z=0$ and
interacting with a single valence-band hole ($h$) confined to the 
plane $z=d$.
This model is appropriate for systems in which the hole concentration 
is very small compared to the electron concentration so that the 
interaction between the holes is negligible.

There are three nearly distinct regions for the interaction of the 
hole and the electron system, which we refer to as weak, strong, 
and intermediate coupling.
In the weak-coupling region ($d$ much larger than the magnetic 
length $\lambda$), the electron--hole interaction is a weak 
perturbation on the eigenstates of the interacting electrons
(Chen and Quinn 1993, 1994a, 1995).
In the strong-coupling region ($d<\lambda$), the hole binds one 
or two electrons to form a neutral ($X$) or negatively charged 
($X^-$) exciton
(
Kheng {\sl et al.}\ 1993,
Buhmann {\sl et al.}\ 1995, 
Shields {\sl et al.}\ 1995,
Finkelstein {\sl et al.}\ 1995, 1996, 
Hayne {\sl et al.}\ 1995, 1999,
Nickel {\sl et al.}\ 1998,
Tischler {\sl et al.}\ 1999,
Wojtowicz  {\sl et al.}\ 1999,
Kim {\sl et al.}\ 2000,
Munteanu {\sl et al.}\ 2000,
W\'ojs and Hawrylak 1995,
Palacios {\sl et al.}\ 1996,
Whittaker and Shields 1997,
W\'ojs {\sl et al.}\ 1998, 1999a, 1999b, 2000a, 2000b).
When $d$ is equal to zero, the neutral exciton is completely uncoupled 
from the remaining $N-1$ electrons due to the ``hidden symmetry''
(Lerner and Lozovik 1981, Dzyubenko and Lozovik 1983, MacDonald 
and Rezayi 1990), and it is only weakly coupled at $0<d<\lambda$.
The $X^-$ is a negatively charged fermion with a similar degenerate 
Landau level (LL) structure to that of an electron (W\'ojs and 
Hawrylak 1995, W\'ojs {\sl et al.}\ 1998, 1999a).
It has Laughlin-like correlations with the remaining $N-2$ electrons 
that can be described by a generalized composite fermion (CF) model
(W\'ojs {\sl et al.}\ 1999b).
In the intermediate-coupling region ($\lambda\leq d\leq2\lambda$), 
the hole can no longer bind a full electron to form an $X$; 
however it can bind one or more Laughlin quasielectrons (QE) to 
form fractionally charged excitons, so-called FCX's (W\'ojs and 
Quinn 2001a, 2001b).
We denote a complex consisting of $n$ QE's bound to a hole by the 
symbol $h$QE$_n$. 
To understand the stability of the $h$QE$_n$ state, it is necessary 
to know the pseudopotentials describing the interactions of a QE--QE 
pair and of a $h$--QE pair as a function of the pair angular momentum
(W\'ojs and Quinn 1998, 1999a, 1999b, 2000a, Quinn and W\'ojs 2000).
In order to determine these pseudopotentials, as well as other 
properties of bound complexes, we have performed exact (within the 
lowest LL) numerical diagonalizations for a nine-electron--one-hole
system as a function of the layer separation $d$ and the magnetic 
field. 
The calculations are performed in Haldane's spherical geometry
(Haldane 1983, Fano {\sl et al.}\ 1986) with the electron--hole 
interaction modeled by $V_{eh}(r)=e^2/\sqrt{r^2+d^2}$, for values 
of $d$ satisfying $0\leq d\leq5\lambda$.

\section{Many-Electron System}
\begin{figure}[t]
\epsfxsize=5in
\epsffile{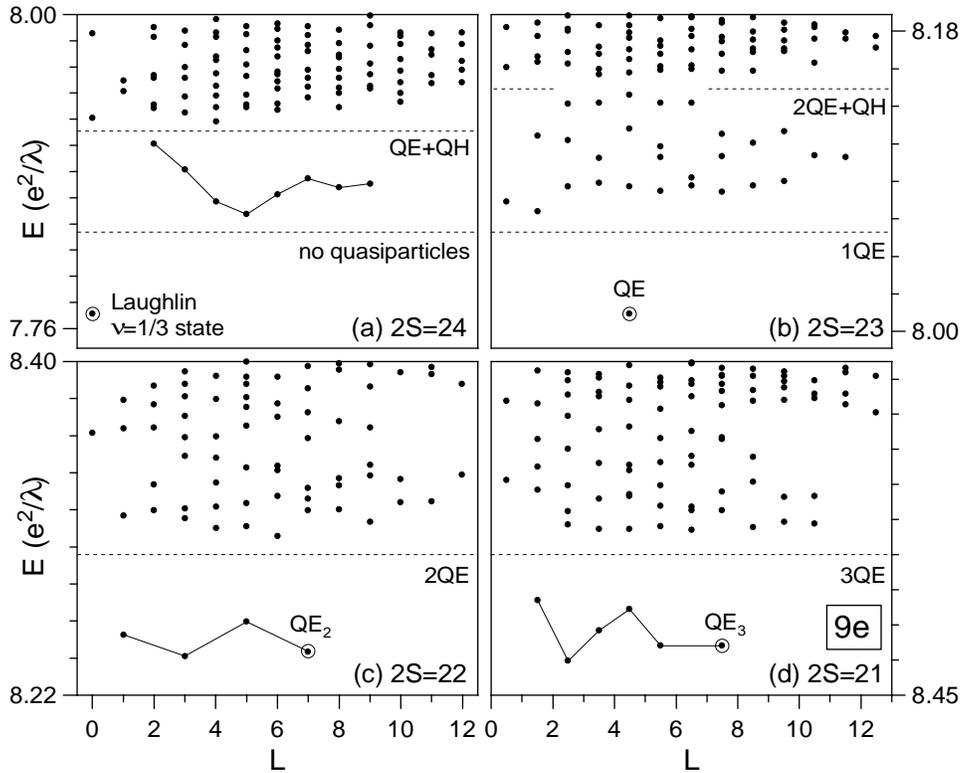}
\caption{
   The energy spectra (energy $E$ vs.\ angular momentum $L$) 
   of the nine-electron system on Haldane's sphere 
   at the monopole strength between $2S=24$ and 21.
   $\lambda$ is the magnetic length.}
\label{fig1}
\end{figure}

To interpret the weak-coupling regime we must begin with the 
understanding of an electron system in the absence of the hole.
In figure~\ref{fig1}(a), (b), (c), and (d) we display the low-energy
spectra of nine electrons at the LL degeneracy $g=2S+1$ corresponding 
to the magnetic monopole strength of $2S=24$, 23, 22, and 21, 
respectively (on Haldane's sphere, the lowest LL is represented by 
a degenerate multiplet at angular momentum $l=S$).
The lowest energy states contain zero (a), one (b), two (c), and 
three (d) QE's in the Laughlin $\nu={1\over3}$ state, and can be 
simply understood using the CF picture.
The effective monopole strength (Chen and Quinn 1994c, W\'ojs and 
Quinn 1998, 1999a, 1999b, 2000a, Quinn and W\'ojs 2000) seen by one 
CF is given by $2S^*=2S-2(N-1)$, and the angular momentum of the 
$k$th CF shell ($k=0$, 1, \dots) is $l_k^*=S^*+k$.
The QE's are the CF's in the first excited shell ($n=1$) and thus 
have $l_{\rm QE}=l_1^*$.
The quasiholes (QH) are the empty states in the lowest CF shell 
($n=0$) and thus have $l_{\rm QH}=l_0^*$.
For $2S=24$ the nine CF's fill completely the $l_0^*=4$ shell giving 
a total angular momentum $L_e=0$.
This non-degenerate ground state (GS) is the Laughlin incompressible 
$\nu={1\over3}$ state.
For $2S=23$, eight of the CF's fill the $l_0^*={7\over2}$ shell, 
and the ninth is a QE with $l_{\rm QE}={9\over2}$.
This gives a total angular momentum $L_e={9\over2}$ for the 
nine-electron GS. 
For $2S=22$ we obtain two QE's each with $l_{\rm QE}=4$; adding the 
angular momenta of these two identical fermions gives $L_e=1\oplus3
\oplus5\oplus7$ as the low-energy band of states.
For $2S=21$ there are three QE's each with $l_{\rm QE}={7\over2}$.
Adding their three angular momenta gives the low-lying nine-electron 
states at $L_e={3\over2}\oplus{5\over2}\oplus{7\over2}\oplus{9\over2}
\oplus{11\over2}\oplus{15\over2}$.

In the absence of the QE--QE interaction (defined by a pseudopotential
$V_{{\rm QE}-{\rm QE}}(L)$, i.e.\ pair interaction energy as 
a function of pair angular momentum), all of the 2QE and 3QE states 
would be degenerate (Chen and Quinn 1994c, Sitko {\sl et al.}\ 1996).
However, the interaction between (charged) QE's exists and removes 
this degeneracy.
In figure~\ref{fig1}(c), the 2QE states with $L=3$ and 7 are lowered 
relative to those with $L=1$ and 5.
For the 3QE states in figure~\ref{fig1}(d), the multiplet at 
$L={5\over2}$ has the lowest energy, and those at $L={3\over2}$ and 
$L={9\over2}$ have the highest energy.
Remarkably, the 2QE and 3QE ``molecule'' states with the maximum 
allowed angular momentum ($L=2l_{\rm QE}-1=7$ and $3l_{\rm QE}-3
={15\over2}$, respectively), that is with the minimum average 
QE--QE separation, have low energy (W\'ojs and Quinn 2000b).
We call these states QE$_2$ and QE$_3$.

All of the many-electron states in the lowest band can be understood 
on the basis of the CF picture; excellent agreement with the numerical 
results is obtained when interactions between quasiparticles (QP$=$QE 
or QH) are included phenomenologically (Sitko {\sl et al.}\ 1996).
In many cases also the first excited band of states can be identified 
using the CF picture.
In this band, one of the CF's is excited to the next higher shell.
Depending on whether a CF from the $n=0$ or 1 shell is excited,
this corresponds to the creation of an additional QE--QH pair or
to the excitation of a QE with $l_{\rm QE}=l_1^*$ to the QE* state 
with $l_{{\rm QE}^*}=l_2^*$.
Let us identify the first excited bands in figure~\ref{fig1}.

For $2S=24$, no QP's occur in the lowest band, and thus the first 
excited band contains a single QE--QH pair.
Because $l_{\rm QH}={1\over2}(N-1)=4$ and $l_{\rm QE}={1\over2}(N+1)
=5$, the multiplets from $L=l_{\rm QE}-l_{\rm QH}=1$ to $l_{\rm QE}
+l_{\rm QH}=N=9$ are expected for such pair.
It is known (Sitko {\sl et al.}\ 1996) that the state at $L=1$ (i.e., 
at the smallest average QE--QH separation) is pushed to higher energy 
(or forbidden) by the strong hard-core QE--QH repulsion.
As a result, the ``magnetoroton'' QE--QH band extends from $L=2$ 
to $N$.

For $2S=23$, two different QP configurations occur in the first 
excited band.
The first consists of a single QE* with $l_{{\rm QE}^*}={11\over2}$ 
giving also the total angular momentum of $L={11\over2}$.
The second consists of a single QH with $l_{\rm QH}={7\over2}$ and 
a pair of QE's each with $l_{\rm QE}={9\over2}$.
Totally ignoring interactions between these QP's gives the following 
set of degenerate angular momentum multiplets for the 2QE$+$QH 
configuration: $L={1\over2}\oplus({3\over2})^2\oplus({5\over2})^3
\oplus({7\over2})^4\oplus({9\over2})^4\oplus({11\over2})^4\oplus
({13\over2})^3\oplus({15\over2})^3\oplus({17\over2})^2\oplus
({19\over2})^2\oplus{21\over2}\oplus{23\over2}$.
The QP--QP interactions (particularly, the QE--QH hard-core repulsion)
will remove the degeneracy of this band and push some of the multiplets 
into the continuum of higher energy states.
However, almost all of the predicted multiplets appear in the numerical 
spectrum in figure~\ref{fig1}(b).

For $2S=22$ there are again two possible configurations for the 
first excited band.
The first contains one QE with $l_{\rm QE}=4$ and one QE* with 
$l_{{\rm QE}^*}=5$, and gives a band of multiplets extending from 
$L=1$ to 9.
The second configuration contains three QE's each with $l_{\rm QE}=4$ 
and a QH with $l_{\rm QH}=3$.
Neglecting QP--QP interactions, these two QP configurations would
yield a degenerate band of multiplets at $L=0^2\oplus1^3\oplus2^5
\oplus3^6\oplus5^4\oplus6^7\oplus7^5\oplus8^4\oplus9^3\oplus10^2
\oplus11\oplus12$.
The numerical spectrum shown in figure~\ref{fig1}(c) contains many 
of these states in a first excited band that is rather well separated 
from the continuum of higher states for $L<3$ and $L>6$, but not 
well defined between these regions.

Finally, for $2S=21$ there are two configurations for the first 
excited band: first with two QE's each with $l_{\rm QE}={7\over2}$ 
and one QE* with $l_{{\rm QE}^*}={9\over2}$, and second with four 
QE's with $l_{\rm QE}={7\over2}$ and one QH with $l_{\rm QH}={5\over2}$.
The former configuration yields the set of multiplets at $L={1\over2}
\oplus({3\over2})^2\oplus({5\over2})^3\oplus({7\over2})^3\oplus
({9\over2})^4\oplus({11\over2})^3\oplus({13\over2})^3\oplus
({15\over2})^2\oplus({17\over2})^2\oplus{19\over2}\oplus{21\over2}$.
The latter one gives $L={1\over2}\oplus({3\over2})^3\oplus
({5\over2})^5\oplus({7\over2})^5\oplus({9\over2})^5\oplus({11\over2})^5
\oplus({13\over2})^5\oplus({15\over2})^3\oplus({17\over2})^2\oplus
{19\over2}\oplus{21\over2}$.
Again, we expect many of these multiplets to be pushed into 
the higher energy continuum by the QE--QH hard-core repulsion.
From the numerical results in figure~\ref{fig1}(d), it can be 
seen that the first excited band is rather well defined for 
$L\le{3\over2}$ and for $L\ge{13\over2}$, where the following 
multiplets occur: $L={1\over2}\oplus({3\over2})^3$ and 
$L=({13\over2})^6\oplus({15\over2})^4\oplus({17\over2})^3\oplus
({19\over2})^2\oplus({21\over2})^2$.

\section{Weak-Coupling Regime}
\begin{figure}[t]
\epsfxsize=5in
\epsffile{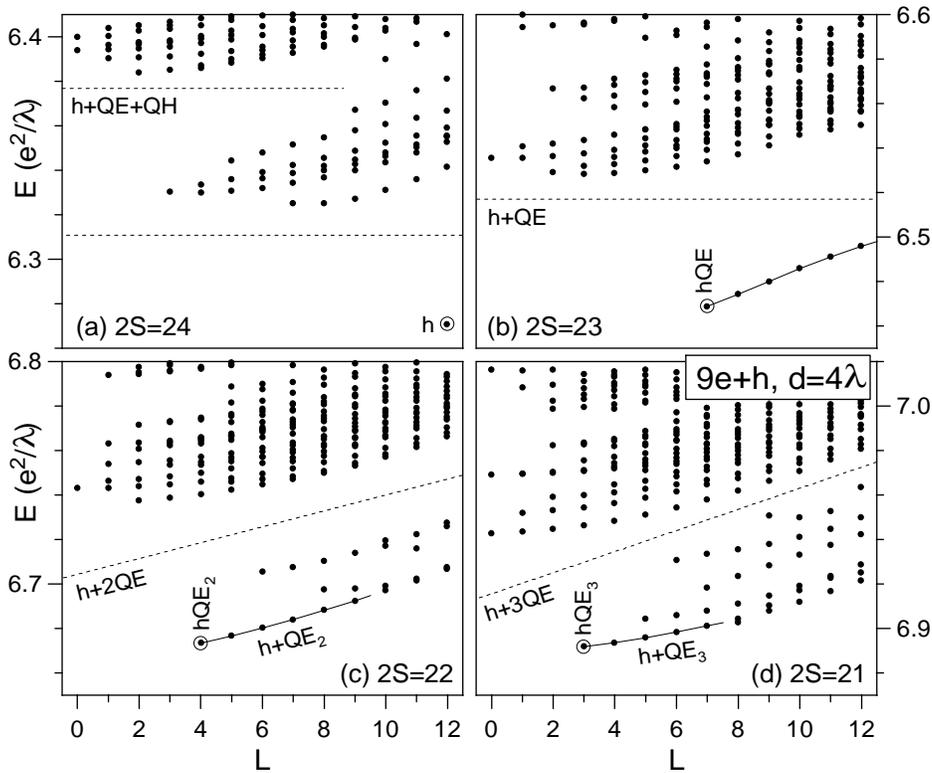}
\caption{
   The energy spectra (energy $E$ vs.\ angular momentum $L$) 
   of the nine-electron--one-hole system on Haldane's sphere 
   at the monopole strength between $2S=24$ and 21.
   The separation of electron and hole layers is $d=4\lambda$
   (weak-coupling regime).
   $\lambda$ is the magnetic length.}
\label{fig2}
\end{figure}

In the weak-coupling limit we expect to obtain fairly well defined 
bands for the electron--hole system by treating the interaction of 
the hole with the electrons as a small perturbation (Chen and Quinn 
1993, 1994a, 1995).
The low-energy bands are clearly visible in figure~\ref{fig2}, and 
they are easily understood on the basis of figure~\ref{fig1} with 
the addition of the angular momenta of the low-energy electron 
multiplets to that of the hole.
The angular momentum of the hole, $l_h$, is equal to $S$, and so 
$l_h=12$, ${23\over2}$, 11, and ${21\over2}$ in figure~\ref{fig1}(a), 
(b), (c), and (d), respectively.
In figure~\ref{fig1}(a) the only electron multiplet in the low-energy 
sector is the Laughlin GS at $L_e=0$.
Therefore, the low-energy band in figure~\ref{fig2}(a) consists of 
a single multiplet at $L=l_h=12$.
In the other frames of figure~\ref{fig1}, the allowed low-energy 
electron multiplets are those of one (b), two (c), and three (d) QE's.
When $l_h$ is added to each of the values of $L_e$ corresponding to 
a low-energy electron state, a band of multiplets is obtained with 
values of $L$ satisfying $|l_h-L_e|\leq L\leq l_h+L_e$.
If the separation $d$ between electron and hole layers were infinite,
so that $V_{eh}=0$, every multiplet in a given band would be degenerate 
and have an energy equal to the energy of the appropriate electron 
state plus the (cyclotron) energy of the hole, which is a constant.
The finite electron--hole interaction causes finite attraction 
between the hole and (negatively charged) QE's so that the energies 
within each band increase with increasing $L$ (decreasing average
$h$--QE separation).
For the electronic multiplets that are close in energy, the $h$--QE 
attraction can also cause mixing or even reversing of the order of 
the corresponding electron--hole bands.

This very simply explains all of the electron--hole bands appearing 
in the low-energy sector of figure~\ref{fig2}.
For example, in figure~\ref{fig2}(c) there are four $h+2$QE bands 
starting at $L=4$, 8, 6, and 10 resulting from the low-energy electronic 
states in figure~\ref{fig1}(c) at $L_e=7$, 3, 5, and 1, respectively.
The bands starting at $L=4$ and 8 have lower energy than those 
starting at $L=6$ and 10 because the electronic multiplets at $L_e=7$ 
and 3 have lower energy than those at $L_e=5$ and 1.
For the first excited sector in figure~\ref{fig2}, we do exactly the 
same thing, add $l_h$ to the allowed values of $L_e$ in the first 
excited sector of figure~\ref{fig1}.
For $2S=24$ this gives bands beginning at $L=3$, 4, 5, \dots, 10, 
corresponding to $L_e=9$, 8, 7, \dots, 2.
All these bands are observed in figure~\ref{fig2}(a).
At $2S=23$, there are a very large number of electronic multiplets 
in the first excited sector of figure~\ref{fig1}(b), but if we 
concentrate on the low-energy multiplets at $L_e\approx l_h=S$ that 
would give low-energy electron--hole bands beginning at low values 
of $L=|l_h-L_e|$, we would expect only a single band beginning at 
$L=0$ and a single band beginning at $L=1$, originating from the 
2QE$+$QH multiplets at $L_e={23\over2}$ and ${21\over2}$, respectively.
Beginning at $L=2$, we would expect two new bands, one at lower and 
one at higher energy, originating from two 2QE$+$QH multiplets at
$L_e={19\over2}$.
We would also expect one low- and one high-energy band beginning 
at $L=3$ arising from the two multiplets at $L_e={17\over2}$, etc.
All these bands can be identified at low $L$ in figure~\ref{fig2}(b).
For $2S=22$ and 21, the first excited bands beginning at low values 
of $L$ can be understood in exactly the same way, that is by picking 
out the low-energy electronic states at values of $L_e$ close to $l_h$.
Thus, all the low-energy electron--hole bands in figure~\ref{fig2}
can be rather well understood in the weak-coupling limit.

Although the attraction between the hole and a given $N$-electron
eigenstate vanishes in the $d\rightarrow\infty$ limit, the most 
tightly bound states of a hole and one, two, and three QE's can 
be identified in figure~\ref{fig2}.
In these states, denoted as $h$QE, $h$QE$_2$, and $h$QE$_3$, a QE 
or an appropriate QE molecule moves as close to the hole as possible.
These states have the smallest $L$ within their $h+$QE, $h+$QE$_2$, 
or $h+$QE$_3$ bands and, together with the ``uncoupled-hole'' state
$h$ in figure~\ref{fig2}(a), are the elementary excitations of the 
weak-coupling regime at finite $d$.

\section{Strong-Coupling Regime}
In the strong-coupling regime the hole binds one or two electrons 
to form an $X$ or an $X^-$.
For $d=0$, the $X$ is totally uncoupled from the remaining $N-1$
electrons (Lerner and Lozovik 1981, Dzyubenko and Lozovik 1983, 
MacDonald and Rezayi 1990), and it is only weakly coupled if $d$ 
is small compared to $\lambda$.
In contrast, the $X^-$ is a negatively charged fermion with LL 
structure just like an electron (W\'ojs and Hawrylak 1995, W\'ojs 
{\sl et al.}\ 1998, 1999a).
Because the $e$--$X^-$ pseudopotential rises more quickly with pair 
angular momentum than the harmonic pseudopotential (W\'ojs and Quinn 
1998, 1999a, 1999b, 2000a, Quinn and W\'ojs 2000) the low-energy states 
of the $X^-$ and $N_e=N-2$ remaining electrons can be described by 
the generalized CF picture (W\'ojs {\sl et al.}\ 1999b) which accounts 
for Laughlin-like correlations among the two types of constituent 
charged particles.

For the system containing one $X^-$ and $N_e$ remaining electrons, 
the effective monopole strengths seen by an electron is given by
(W\'ojs {\sl et al.}\ 1999b)
\begin{equation}
   2S_e^*=2S-\left(m_e-1\right)\left(N_e-1\right)-m_{eX^-},
\label{eq1}
\end{equation}
while that seen by an $X^-$ is
\begin{equation}
   2S_{X^-}^*=2S-m_{eX^-}N_e.
\label{eq2}
\end{equation}
Here $m_{eX^-}$ is the exponent describing the Laughlin correlations 
between the $X^-$ and each electron in the two-component many-body 
wavefunction.
In the generalized CF picture, electrons and $X^-$'s are converted
into two types of CF's: CF-$e$ and CF-$X^-$.
The angular momenta of their lowest shells are $l_e^*=S_e^*$ and 
$l_{X^-}^*=S_{X^-}^*-1$.
The following types of Laughlin QP's can be defined:
the particles in the first excited CF-$e$ shell are ``$e$-type'' 
quasielectrons (QE-$e$) with $l_{{\rm QE}e}=l_e^*+1$, the empty 
states in the lowest CF-$e$ shell are ``$e$-type'' quasiholes 
(QH-$e$) with $l_{{\rm QH}e}=l_e^*$, and a single particle in 
the lowest CF-$X^-$ shell is an ``$X^-$-type'' quasielectron 
(QE-$X^-$) with $l_{{\rm QE}X^-}=l_{X^-}^*$.
For simplicity, in the figures we denote QH-$e$ and QE-$X^-$ 
by QH and $X^-$, respectively.

\begin{figure}[t]
\epsfxsize=5in
\epsffile{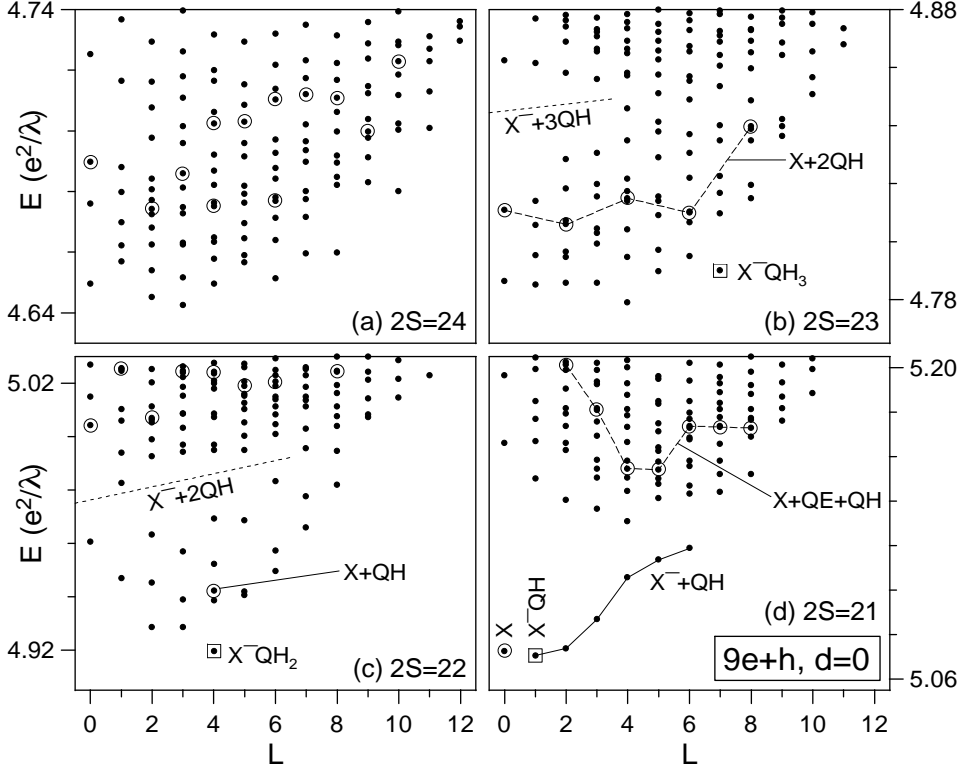}
\caption{
   The energy spectra (energy $E$ vs.\ angular momentum $L$) 
   of the nine-electron--one-hole system on Haldane's sphere 
   at the monopole strength between $2S=24$ and 21.
   The separation of electron and hole layers is $d=0$
   (strong-coupling regime).
   $\lambda$ is the magnetic length.}
\label{fig3}
\end{figure}

Let us turn to the nine-electron--one-hole spectra shown in 
figure~\ref{fig3}.
For $2S=21$, the energies of the multiplicative states (containing 
one uncoupled $X$) are exactly the same as those of eight electrons, 
with the energy shifted by the $X$ binding energy.
For the eight electrons, the lowest CF shell has $l_0^*={7\over2}$, 
and it is filled completely giving an $L=0$ Laughlin GS. 
An excitation of one CF to the next shell gives a ``magnetoroton'' 
QE--QH band extending from $L=2$ to 8, marked with a dashed line.
The lowest non-multiplicative states contain seven electrons and 
an $X^-$.
For $m_e=3$ and $m_{eX^-}=2$ we obtain $2S_e^*=2S_{X^-}^*=7$.
The eight states in the lowest CF-$e$ shell contain one QH-$e$ 
with $l_{{\rm QH}e}={7\over2}$, and the single QE-$X^-$ has 
$l_{{\rm QE}X^-}={5\over2}$.
The addition of $l_{{\rm QH}e}$ and $l_{{\rm QE}X^-}$ gives the 
band of multiplets extending from $L=1$ to 6.
These states are the bound states of a QH-$e$--QE-$X^-$ pair, 
connected by a solid line in figure~\ref{fig3}(d).
Their energy increases with increasing $L$, just as that of the 
neutral exciton does.
Our interpretation of this low-lying band of non-multiplicative 
states is totally different from that of a neutral exciton with 
finite momentum which is dressed by magnetorotons of the Laughlin 
condensed state, suggested by previous authors (Apalkov and Rashba 
1992, 1993, MacDonald {\sl et al.}\ 1992, Wang {\sl et al.}\ 1992).
The latter picture does not work because the coupling of an exciton 
with a finite electric dipole moment to the electrons is too strong 
to be treated perturbatively.

For $2S=22$, there is a single low-energy multiplet at $L=4$ that 
is a multiplicative state.
It corresponds to a single QH in the $l_0^*$ shell.
The non-multiplicative states exhibit a low-energy band containing 
the multiplets at $L=0\oplus1\oplus2^3\oplus3^3\oplus4^4\oplus5^3
\oplus6^3\oplus7^2\oplus8^2\oplus9\oplus10$.
These arise from two QH-$e$'s each with $l_{{\rm QH}e}=4$ plus one 
QE-$X^-$ with $l_{{\rm QE}X^-}=3$.
The GS called $X^-$QH$_2$ marked with an open square is the most 
tightly bound state of QE-$X^-$ and a (QH-$e$)$_2$ molecule.
Its angular momentum $L=4$ results from adding two $l_{{\rm QH}e}$'s
to obtain $l_{({\rm QH}e)_2}=2l_{{\rm QH}e}-1=7$, and then adding 
to it one $l_{{\rm QE}X^-}$ to obtain $l_{X^-{\rm QH}_2}
=|l_{{\rm QE}X^-}-l_{({\rm QH}e)_2}|=4$.

For $2S=23$, the low-energy band of multiplicative states contains 
two QH's each with $l_{\rm QH}={9\over2}$, resulting in the multiplets 
$L=0\oplus2\oplus4\oplus6\oplus8$ (connected with a dashed line).
The non-multiplicative states have a low-energy band containing three 
QH-$e$'s each with $l_{{\rm QH}e}={9\over2}$ and one QE-$X^-$ with 
$l_{{\rm QE}X^-}={7\over2}$.
Addition of these angular momenta gives a band of multiplets at 
$L=0\oplus1^4\oplus2^6\oplus3^7\oplus4^8\oplus5^9\oplus6^8\oplus7^8
\oplus8^7\oplus9^5\oplus10^4\oplus11^3\oplus12^2\oplus13\oplus14$.
The angular momentum of the most tightly bound state of an QE-$X^-$ 
and a (QH-$e$)$_3$ molecule is $l_{X^-{\rm QH}_3}=7$.
This state is (most likely) the one marked with an open rectangle.

Finally, for $2S=24$ the low-lying band of multiplicative state 
contains $L=0\oplus2\oplus3\oplus4^2\oplus5\oplus6^2\oplus7\oplus8
\oplus9\oplus10\oplus12$, arising from three QH's each with 
$l_{\rm QH}=5$.
The lowest non-multiplicative band contains four QH-$e$'s each with 
$l_{{\rm QH}e}=5$ and one QE-$X^-$ with $l_{{\rm QH}X^-}=4$.
It produces 194 multiplets starting with $L=0^3\oplus1^6\oplus\dots$ 
and ending with $15^3\oplus16^2\oplus17\oplus18$.
Only the lower-energy multiplets of this set are shown in 
figure~\ref{fig3}(d) in which we have restricted the range of 
energy and $L$.

It is really quite remarkable that the myriad of multiplets 
associated with the lowest band of both multiplicative and 
non-multiplicative states appear in the numerical spectra shown 
in figure~\ref{fig3} exactly as predicted by the generalized CF 
picture.
This simple model ignores the interaction between QP's which can 
lead to some overlap of the lowest bands with higher bands at some 
values of $L$ when the QP--QP interaction energy becomes large.

\section{Intermediate-Coupling Regime}
When the separation $d$ increases beyond roughly one magnetic length 
$\lambda$, the attractive Coulomb potential of the hole is not strong 
enough (and its resolution is not high enough) to bind a full electron. 
We know this from numerical calculations for a single electron and 
hole as a function of $d$ at different values of $2S$.
When many electrons are present, it is possible that the $X$ 
and even the $X^-$ may persist to larger values of $d$ due to 
correlations within the entire electron system.
However, from knowing that at large values of $d$, $V_{eh}$ acts 
as a small perturbation on the $N$-electron eigenstates, we might 
guess that Laughlin QE's rather than ``full'' electrons could 
bind to the hole to form FCX's.
The prediction that QE's can bind to a hole at $d>\lambda$ at 
which ``full'' electrons do not is based on two simple facts:
(i) negative electric charge of QE's ($q_{\rm QE}=-{1\over3}e$ 
at $\nu={1\over3}$) causes $h$--QE attraction, and
(ii) because the number of QE's is smaller than the number of 
electrons, their characteristic separation is larger and their 
interaction energy is smaller than those of electrons, and thus 
smaller strength and lower resolution of the hole's attractive 
potential are sufficient to pick one out of many interacting QE's 
to form a bound state.

\begin{figure}[t]
\epsfxsize=5in
\epsffile{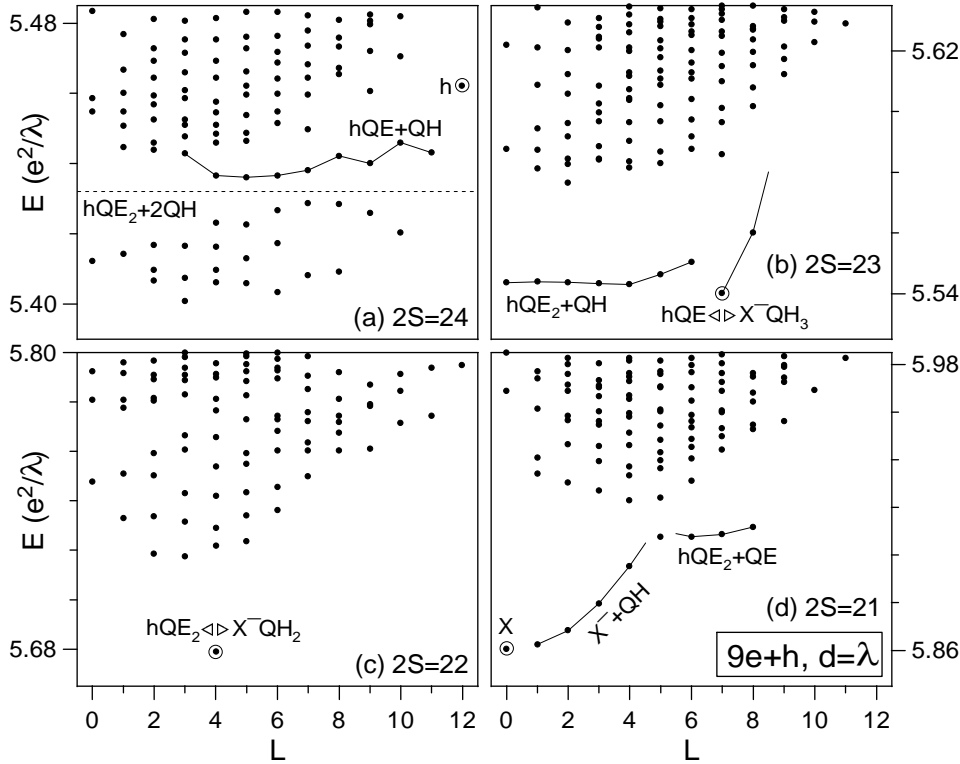}
\caption{
   The energy spectra (energy $E$ vs.\ angular momentum $L$) 
   of the nine-electron--one-hole system on Haldane's sphere 
   at the monopole strength between $2S=24$ and 21.
   The separation of electron and hole layers is $d=\lambda$
   (intermediate-coupling regime).
   $\lambda$ is the magnetic length.}
\label{fig4}
\end{figure}

\begin{figure}[t]
\epsfxsize=5in
\epsffile{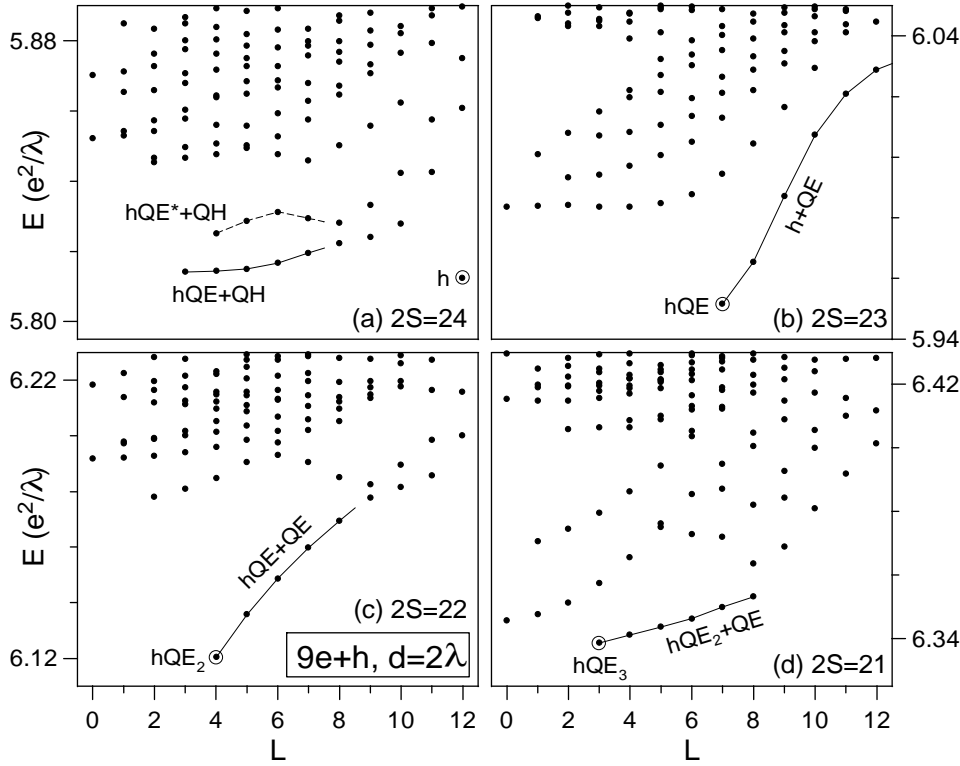}
\caption{
   The energy spectra (energy $E$ vs.\ angular momentum $L$) 
   of the nine-electron--one-hole system on Haldane's sphere 
   at the monopole strength between $2S=24$ and 21.
   The separation of electron and hole layers is $d=2\lambda$
   (intermediate-coupling regime).
   $\lambda$ is the magnetic length.}
\label{fig5}
\end{figure}

In figures~\ref{fig4} and \ref{fig5} we present the spectra obtained 
for the nine-electron--one-hole system at $d=\lambda$ and $2\lambda$. 
We shall attempt to interpret the low-lying bands in terms of the
elementary excitations and their interactions.
Because $d=\lambda$ and $2\lambda$ are on the borders of strong- 
and weak-coupling regions, respectively, the interpretation is 
not always unique.
The stable elementary excitations of the strong-coupling regime 
are $X$, $X^-$, $X^-$QH, or $X^-$QH$_2$, and those of the 
weak-coupling regime are $h$, $h$QE, and $h$QE$_2$.
At the values of $2S=21$ and 24, at which the lowest-lying states or 
bands at small and large $d$ occur at different $L$, the $d$-driven 
phase transition between the two regimes is of the first order and 
the crossing of the appropriate energy levels is observed in the 
spectrum.
However, this is not the case for $2S=22$ where $X^-$QH$_2$ has 
the same $L={1\over2}(N-1)$ as $h$QE$_2$, or at $2S=23$ where 
$X^-$QH$_3$ has the same $L=N-2$ as $h$QE.
In these spectra, the anti-crossing of energy levels occurs, and 
the analysis of wavefunctions is needed to detect the phase 
transition (of the second order) between the two regions.

For $2S=21$, it is clear from figures~\ref{fig4}(d) and \ref{fig5}(d) 
that the band of states extending from $L=0$ to 6 appearing in 
figure~\ref{fig3}(d) is still present beyond $d=\lambda$.
For $L=5$ and 6 it crosses into the continuum of higher energy states.
For $d=0$, this band was identified as the multiplicative state of 
eight electrons in a Laughlin incompressible state plus an $X$ in 
its GS at $L=0$, and the $X^-$QH band predicted by the generalized 
CF picture at $L=1$, 2, \dots, 6.
Hence, we conclude that the $X$ and $X^-$ bound states persist beyond 
$d=\lambda$ in this system.
The band of states extending to $L=8$ which crosses the $X^-$QH band 
at $L=5$ can be associated with a bound state of the hole and two 
Laughlin QE's, interacting with a third QE.
The $h$QE$_2$ is the most strongly bound FCX, as we shall see later 
in this section.
Since the Laughlin QE has $l_{\rm QE}=l_1^*=S-(N-1)+1={7\over2}$, 
the allowed values of $L_{2{\rm QE}}$ are $0\oplus2\oplus4\oplus6$.
The 2QE state with smallest average QE--QE separation is the QE$_2$ 
molecule with $l_{{\rm QE}_2}=6$, and the $h$QE$_2$ state has 
$l_{h{\rm QE}_2}=|l_h-l_{{\rm QE}_2}|={9\over2}$.
Adding to $l_{h{\rm QE}_2}$ the angular momentum of the third 
QE as if it were distinguishable from the two QE's within the 
$h$QE$_2$ would give a band of states extending from 
$L=|l_{h{\rm QE}_2}-l_{\rm QE}|=1$ to $l_{h{\rm QE}_2}+l_{\rm QE}=8$.
However, the three QE's in the $h$QE$_2+$QE band are identical 
fermions and must obey and the Pauli exclusion principle.
The exclusion principle forbids a number of $h$QE$_2+$QE pair 
states corresponding to the smallest $h$QE$_2$--QE separation.
The forbidden states can be most easily identified by noticing 
that the largest angular momentum of three QE's is $l_{{\rm QE}_3}
=3l_{\rm QE}-3={15\over2}$ which, when added to $l_h={21\over2}$, 
cannot result in a total $h+3$QE angular momentum smaller than 3.
Thus, the $L=1$ and 2 states of the $h$QE$_2$--QE pair are 
forbidden and the $h$QE$_2$--QE band is expected to extend from
$L=3$ to 8, exactly as observed in figure~\ref{fig5}(d).
Remarkably, although the $h$QE$_2$ and QE have opposite charges 
($q_{h{\rm QE}_2}=+{1\over3}e$ and $q_{\rm QE}=-{1\over3}e$), the 
interaction energy does not increase with increasing $L$ as it 
does for the electron--hole pair.
This is due to the complex (not point particle) structure of the 
constituents, which makes the $h$QE$_2$--QE interaction not generally 
attractive (and is responsible for the instability of the $h$QE$_3$ 
complex).
In figure~\ref{fig5}(d), the entire $h$QE$_2$ band at $L=3$ to 8 is
observed at $d=2\lambda$, as it is lower in energy that the remnants 
of the $X^-$--QH band.

For $2S=22$, a second-order transition between the $X^-$QH$_2$ GS 
of the strong coupling at $L=4$ and the $h$QE$_2$ GS of the weak 
coupling at the same $L$ is observed in figures~\ref{fig4}(c) and 
\ref{fig5}(c).
Although it is not clear from the energy spectrum alone, the two
GS's anti-cross at $d\approx\lambda$.
The low-lying excitations of the $h$QE$_2$ state at $d=2\lambda$ 
are connected with a solid line in figure~\ref{fig5}(c).
At this intermediate value of $d$, they are best interpreted as 
a dispersion of a $h$QE--QE pair.
This changes at larger $d$ when the QE--QE interaction becomes
dominant over the $h$QE--QE interaction and the low-lying 
excitations of the $h$QE$_2$ state turn into the $h$--QE$_2$ pair
excitations identified in figure~\ref{fig2}(c).

For $2S=23$, a similar second-order transition between the $X^-$QH$_3$
and $h$QE states occurs at $L=7$.
Additionally, a well-defined band of $h$QE$_2$--QH pair states occurs
at $L=0$ to 6.
This band occurs only in the intermediate-coupling region 
(at $d\approx\lambda$) and cannot be found in figure~\ref{fig2} 
or \ref{fig3}.
The range of $d$ in which it has low energy is determined by the
competition between the $h$QE--QE binding energy gained through the 
formation of a $h$QE$_2$ state and the Laughlin energy gap to create
an additional QE--QH pair.
The angular momenta of the $h$QE$_2$--QH states result from adding 
$l_{h{\rm QE}_2}={7\over2}$ to $l_{\rm QH}={7\over2}$ to obtain
$L=0$ to 7.
The $L=7$ state corresponding to the smallest average $h$QE$_2$--QH 
separation is most likely pushed to a high energy because of the 
QE--QH hard-core.

Finally, at 2S=24 the lowering of energy of the $h$ state at $L=12$
with increasing $d$ is observed.
In this state the hole becomes completely uncoupled from the 
nine-electron Laughlin GS in the $d\rightarrow\infty$ limit.
Other low-energy bands that occur in the intermediate-coupling 
regime involve one or two QE--QH pairs spontaneusly induced in 
the electron system to screen the charge of the hole.
At a smaller $d=\lambda$, the coupling of the hole to the Laughlin 
excitations of the electron GS is stronger and the $h$QE$_2+$2QH 
band has the lowest energy.
At a larger $d=2\lambda$, the coupling is weaker (and so is the 
$h$--QE attraction compared to the Laughlin gap) and the band 
of $h$QE--QH pair states at $L=3$ to 11 moves down in energy 
relative to the $h$QE$_2+$2QH band.
The band starting at $L=4$ and going slightly above the $h$QE--QH 
band is best interpreted as the band of $h$QE*--QH states in which 
$h$QE* is the first excited state of $h$QE, with the angular 
momentum $l_{h{\rm QE}^*}=l_{h{\rm QE}}+1$.
As shown in figure~\ref{fig2}(a), at even higher $d$, when the distant 
hole uncouples from the electron system, all bands involving QE--QH
pairs reconstruct and the $h$ GS becomes stable.
The angular momenta of the $h$QE--QH and $h$QE*--QH pairs can be
calculated by adding $l_{h{\rm QE}}=7$ or $l_{h{\rm QE}^*}=8$ to 
$l_{\rm QH}=4$ to obtain $L=3$ to 11 and $L=4$ to 12, respectievely.
For the $h$QE$_2+$2QH band, $l_{h{\rm QE}_2}=3$ must be added to
all possible values of $L_{2{\rm QH}}=1\oplus3\oplus5\oplus7$.
The result is $L=0\oplus1\oplus2^3\oplus3^3\oplus4^4\oplus5^3\oplus
6^3\oplus7^2\oplus8^2\oplus9\oplus10$; all these multiplets occur
below the dotted line in figure~\ref{fig4}(a).

\section{Photoluminescence}
In the preceding sections we have identified the low-lying elementary 
excitations of an ideal system containing $N$ electrons confined to 
a plane and one hole confined to a neighboring parallel plane in the 
high magnetic field limit.
In this limit the cyclotron energy $\hbar\omega_c$ is so large 
compared to the Coulomb energy $e^2/\lambda$ that only the lowest 
LL need be considered (W\'ojs and Quinn 2001a, 2001b).
In actual experiments at finite magnetic fields, the mixing of 
LL's by the Coulomb interaction occurs.
It is particularly important in the strong-coupling regime.
While only one bound $X^-$ state exists in the lowest LL --
the non-radiative triplet (W\'ojs and Hawrylak 1995, Palacios 
{\sl et al.}\ 1996) with parallel electron spins identified in 
figure~\ref{fig3} -- the LL mixing leads to the binding of other 
$X^-$ states, particularly of the optically active singlet
(Buhmann {\sl et al.}\ 1995, Kheng {\sl et al.}\ 1993, Whittaker 
and Shields 1997, W\'ojs {\sl et al.}\ 2000a, 2000b) observed in PL.
The LL mixing is less important for the intermediate and weak coupling.
Real systems are also complicated by finite quantum well widths, 
different barrier heights for electrons and holes, non-parabolicity 
of the energy bands, spin--orbit coupling, etc.

In this section we will discuss PL (radiative recombination of an 
electron--hole pair) in the ``theoretical'' situation in which the 
interacting particles are confined to planes and only the lowest LL 
is taken into account ($B\rightarrow\infty$).
There are two symmetries that limit the possible radiative decay 
processes (W\'ojs and Quinn 2000a, 2000b, 2001a, 2001b).
The most important one is the geometrical symmetry: translational 
invariance on a plane (or rotational invariance on Haldane's sphere).
On the plane there are two conserved orbital quantities ${\cal M}$, 
the $z$-component of angular momentum, and ${\cal K}$, an additional 
quantum number associated with the partial decoupling of the 
center-of-mass motion of the electron--hole system in the magnetic 
field (Avron {\sl et al.}\ 1978, Dzyubenko 2000).
States in a given LL all have the same value of ${\cal L}={\cal M}
+{\cal K}$, and different states in a LL are labeled by ${\cal K}=0$, 
1, 2, \dots.
On the spherical surface, the total angular momentum $L$ and its 
$z$-component $L_z$ are conserved.
The other symmetry is the ``hidden symmetry'' which is exact in 
the lowest LL at  $d=0$ and only weakly broken in the entire 
weak-coupling regime.
The ``hidden symmetry'', that is the particle--hole symmetry of 
the electron--valence-band-hole Hamiltonian $H$, depends on equal 
magnitude of the electron--electron and electron--hole interactions
in the lowest LL.
This symmetry makes the commutator of the PL operator ${\cal P}$ 
which annihilates an optically active electron--hole pair with 
$H$ proportional to ${\cal P}$ itself (Dzyubenko and Lozovik 1983, 
MacDonald and Rezayi 1990).
Because of this, only the multiplicative states (containing one 
uncoupled neutral exciton) are radiative at $d=0$.
At $d>0$, the states originating from other, non-multiplicative 
states become radiative, but their PL intensity remains very low 
at $d<\lambda$.
Thus the PL spectrum in the weak-coupling regime gives information 
about the binding of the $X$, but not about original 
electron--electron correlations in the 2DEG.

For weak and intermediate coupling the system is best described in 
terms of the $h$QE$_n$ bound states (FCX's).
At $\nu\approx{1\over3}$, the recombination process can be thought 
of as (Chen and Quinn 1994b)
\begin{equation}
   h+n{\rm QE}\rightarrow(3-n){\rm QH}+\gamma.
\label{eq3}
\end{equation}
In other words, the hole plus $n=0$, 1, 2, or 3 Laughlin QE's combine 
to give off a photon $\gamma$ plus $3-n=3$, 2, 1, or 0 Laughlin QE's. 
Other processes, involving additional QE--QH pairs, have much smaller
oscillator strength.
In figure~\ref{fig2} we have shown the low-lying bands for the 
nine-electron--one-hole systems that contain zero (a), one (b), 
two (c), and three (d) QE's, respectively.
At zero temperature only the lowest state in each frame is occupied 
and can serve as an initial state.
At finite temperature $T$, the probability of the eigenstate of energy 
$E$ being occupied is proportional to $\exp\left[-E/k_BT\right]$.
The PL spectra are obtained by evaluating the transition rate 
$w_{i\rightarrow f}$ between low-lying initial states $\left|i\right>$ 
of the $N$-electron--one-hole system and final states $\left|f\right>$ 
of the $(N-1)$-electron system,
\begin{equation}
   w_{i\rightarrow f}={\rm const}\cdot|\left<f|{\cal P}|i\right>|^2.
\label{eq4}
\end{equation}
We have used the eigenfunctions of the nine-electron--one-hole system
and of the eight-electron system obtained in numerical diagonalization
to evaluate $w_{i\rightarrow f}$.
We find the following results for weak and intermediate coupling
(W\'ojs and Quinn 2001a, 2001b):

(i) For weak coupling, the PL intensity is weak. 
However, the PL spectra can involve one or more peaks of different 
relative intensities whose energies depend on the value of $n$ in 
Eq.~(\ref{eq3}).

(ii) For intermediate coupling, the strongest emission is that of 
the strongly bound and radiative $h$QE$_2$. 
The recombination of the $h$QE ground state is forbidden by the 
conservation of $L$ (or ${\cal K}$), but the excited state $h$QE*
is radiative.
The ``uncoupled-hole'' state $h$ is radiative and, finally, the 
$h$QE$_3$ ``anyon exciton'' proposed earlier (Rashba and Portnoi
1993) is neither bound nor radiative.

Let us illustrate the $\Delta L=0$ optical selection rule on the 
examples of the $h$QE$_2$ and $h$QE recombination.
An isolated $h$QE$_2$ state occurs in the nine-electron--one-hole 
system at $2S=22$; see figure~\ref{fig5}(c).
Its angular momentum, $l_{h{\rm QE}_2}=4$, arose from $l_{\rm QE}=4$ 
and $l_{{\rm QE}_2}=7$ combined with $l_h=11$.
After the recombination, we are left with one QH in the eight-electron 
system, which has the same angular momentum, $l_{\rm QH}=S^*=4$.
Therefore, $\left|i\right>=\left|h{\rm QE}_2\right>$ and 
$\left|f\right>=\left|{\rm QH}\right>$ each have $L=4$, and the 
optical process is allowed by the $\Delta L=0$ selection rule.

The isolated $h$QE and $h$QE* states occur at $2S=23$; see 
figure~\ref{fig5}(b).
Their angular momenta, $l_{h{\rm QE}}=7$ and $l_{h{\rm QE}^*}=8$, 
resulted from $l_{\rm QE}={9\over2}$ combined with $l_h={23\over2}$.
In the final state, two QH's occur each with $l_{\rm QH}={9\over2}$.
The allowed angular momenta for the 2QH pair states are $L_{2{\rm QH}}
=0\oplus2\oplus4\oplus6\oplus8$.
Comparing angular momenta of initial and final states we obtain 
that the $\left|i\right>=\left|h{\rm QE}\right>$ initial ground 
state cannot recombine to create a 2QH pair, 
$\left|f\right>=\left|2{\rm QH}\right>$.
However, the excited initial state 
$\left|i\right>=\left|h{\rm QE}^*\right>$ is optically active,
and the final state for its recombination is the
$\left|f\right>=\left|{\rm QH}_2\right>$ molecule.
Because $h$QE* is an excited state, its PL line is expected at $T>0$.

\section{Summary}
We have calculated numerically the exact eigenstates of the 
nine-electron--one-hole system on Haldane's sphere in the ``ideal'' 
theoretical limit of $\hbar\omega_c\rightarrow\infty$ and zero 
widths of electron and hole layers.
We have shown how the low-lying bands in strong, weak and intermediate
coupling can be understood in terms of rather simple elementary 
composite particles ($X$, $X^-$, $h$QE, $h$QE$_2$, etc.) and electrons.
We have studied the oscillator strength for radiative recombination
and found that certain radiative decay processes are strongly inhibited.
In the strong-coupling region, only the multiplicative states 
(or, at finite but small values of $d$, the states arising from them) 
have appreciable oscillator strength.
For intermediate and strong coupling the recombination of the $h$QE$_2$ 
bound state has the highest intensity.
The ``uncoupled hole'' $h$ and the excited state $h$QE* are also
radiative, but the recombination of the latter state will only be
observed if this state is occupied at a finite temperature at which 
the PL experiment is performed.

\section{Acknowledgment}
The authors thank Izabela Szlufarska for help on this problem, 
and acknowledge the support of the Materials Research Program 
of Basic Energy Sciences, US Department of Energy. 
AW acknowledges discussions with P. Hawrylak (IMS NRC Ottawa) 
and M. Potemski (HMFL Grenoble), and partial support of KBN 
Grant 2P03B05518.
KSY acknowledges support from Korea Research Foundation.

\section*{References}

\begin{list}{}{
   \small
   \leftmargin 0.5in
   \itemindent -0.5in
   \itemsep -0.02in}

\item 
{\sc Apalkov, V. M.}, and {\sc Rashba, E.}, 
1992, {\sl Phys. Rev.} B, {\bf46}, 1628;
1993, {\sl Phys. Rev.} B, {\bf48}, 18~312.

\item 
{\sc Avron, J. E.}, {\sc Herbst, I. W.}, and {\sc Simon, B.}, 
1978, {\sl Ann. Phys.}, {\bf114}, 431.

\item 
{\sc Brown, S. A.}, {\sc Young, J. F.}, {\sc Brum, J. A.}, 
{\sc Hawrylak, P.}, and {\sc Wasilewski, Z.}, 
1996, {\sl Phys. Rev.} B, {\bf54}, R11~082.

\item 
{\sc Buhmann, H.}, {\sc Joss, W.}, {\sc von Klitzing, K.}, 
{\sc Kukushkin, I. V.}, {\sc Martinez, G.}, {\sc Plaut, A. S.}, 
{\sc Ploog, K.}, and {\sc Timofeev, V. B.}, 
1990, {\sl Phys. Rev. Lett.}, {\bf65}, 1056.

\item 
{\sc Buhmann, H.}, {\sc Joss, W.}, {\sc von Klitzing, K.}, 
{\sc Kukushkin, I. V.}, {\sc Martinez, G.}, {\sc Ploog, K.}, 
and {\sc Timofeev, V. B.}, 
1991, {\sl Phys. Rev. Lett.}, {\bf66}, 926.

\item 
{\sc Buhmann, H.}, {\sc Mansouri, L.}, {\sc Wang, J.}, 
{\sc Beton, P. H.}, {\sc Mori, N.}, {\sc Heini, M.}, and 
{\sc Potemski, M.}, 
1995, {\sl Phys. Rev.} B, {\bf51}, 7969.

\item 
{\sc Chen, X. M.}, and {\sc Quinn, J. J.}, 
1993, {\sl Phys. Rev. Lett.}, {\bf70}, 2130;
1994a, {\sl Phys. Rev.} B, {\bf50}, 2354;
1994b, {\sl Solid St. Commun.}, {\bf90}, 303.
1994c, {\sl Solid St. Commun.}, {\bf92}, 865.
1995, {\sl Phys. Rev.} B, {\bf51}, 5578.

\item 
{\sc Dzyubenko, A. B.}, and {\sc Lozovik, Yu. E.}, 
1983, {\sl Fiz. Tverd. Tela}, {\bf25}, 1519
[1983, {\sl Sov. Phys. Solid State}, {\bf25}, 874].

\item 
{\sc Dzyubenko, A. B.}, 
2000, {\sl Solid St. Commun.}, {\bf113}, 683.
   
\item 
{\sc Fano, G.}, {\sc Ortolani, F.}, and {\sc Colombo, E.}, 
1986, {\sl Phys. Rev.} B, {\bf34}, 2670.

\item 
{\sc Finkelstein, G.}, {\sc Shtrikman, H.}, and 
{\sc Bar-Joseph, I.}, 
1995, {\sl Phys. Rev. Lett.}, {\bf74}, 976;
1996, {\sl Phys. Rev.} B, {\bf53}, R1709.

\item 
{\sc Goldberg, B. B.}, {\sc Heiman, D.}, {\sc Pinczuk, A.}, 
{\sc Pfeiffer, L. N.}, and {\sc West, K.}, 
1990, {\sl Phys. Rev. Lett.}, {\bf65}, 641.

\item 
{\sc Goldys, E. M.}, {\sc Brown, S. A.}, {\sc Davies, A. G.}, 
{\sc Newbury, R.}, {\sc Clark, R. G.}, {\sc Simmonds, P. E.}, 
{\sc Harris, J. J.}, and {\sc Foxon, C. T.}, 
1992, {\sl Phys. Rev.} B, {\bf46}, R7957.

\item 
{\sc Gravier, L.}, {\sc Potemski, M.}, {\sc Hawrylak, P.}, 
and {\sc Etienne, B.}, 
1998, {\sl Phys. Rev. Lett.}, {\bf80}, 3344.

\item 
{\sc Haldane, F. D. M.}, 
1983, {\sl Phys. Rev. Lett.}, {\bf51}, 605.

\item 
{\sc Hayne, M.}, {\sc Jones, C. L.}, {\sc Bogaerts, R.}, 
{\sc Riva, C.}, {\sc Usher, A.}, {\sc Peeters, F. M.}, 
{\sc Herlach, F.}, {\sc Moshchalkov, V. V.}, and {\sc Henini, M.}, 
1999, {\sl Phys. Rev.} B, {\bf59}, 2927.

\item 
{\sc Heiman, D.}, {\sc Goldberg, B. B.}, {\sc Pinczuk, A.}, 
{\sc Tu, C. W.}, {\sc Gossard, A. C.}, and {\sc English, J. H.}, 
1988, {\sl Phys. Rev. Lett.}, {\bf61}, 605.

\item 
{\sc Jiang, Z. X.}, {\sc McCombe, B. D.}, and {\sc Hawrylak, P.}, 
1998, {\sl Phys. Rev. Lett.}, {\bf81}, 3499.

\item 
{\sc Kheng, K.}, {\sc Cox, R. T.}, {\sc d'Aubigne Y. M.}, 
{\sc Bassani, F.}, {\sc Saminadayar, K.}, and {\sc Tatarenko, S.}, 
1993, {\sl Phys. Rev. Lett.}, {\bf71}, 1752.

\item 
{\sc Kim, Y.}, {\sc Munteanu, F. M.}, {\sc Perry, C. H.}, 
{\sc Rickel, D. G.}, {\sc Simmons, J. A.}, and {\sc Reno, J. L.}, 
2000, {\sl Phys. Rev.} B, {\bf61}, 4492;

\item 
{\sc Kukushkin, I. V.}, {\sc Haug, R. J.}, {\sc von Klitzing, K.}, 
{\sc Eberl, K.}, and {\sc T\"otemeyer, K.}, 
1994, {\sl Phys. Rev.} B, {\bf50}, 11~259.

\item 
{\sc Lerner, I. V.}, and {\sc Lozovik, Yu. E.}, 
1981, {\sl Zh. Eksp. Teor. Fiz.}, {\bf80}, 1488
[1981 {\sl Sov. Phys. JETP}, {\bf53}, 763].

\item 
{\sc MacDonald, A. H.}, and {\sc Rezayi, E. H.}, 
1990, {\sl Phys. Rev.} B, {\bf42}, 3224.

\item 
{\sc MacDonald, A. H.}, {\sc Rezayi, E. H.}, and {\sc Keller, D.}, 
1992, {\sl Phys. Rev. Lett.}, {\bf68}, 1939.

\item 
{\sc Nickel, H. A.}, {\sc Herold, G. S.}, {\sc Yeo, T.}, 
{\sc Kioseoglou, G.}, {\sc Jiang, Z. X.}, {\sc McCombe, B. D.}, 
{\sc Petrou, A.}, {\sc Broido, D.}, and {\sc Schaff, W.}, 
1998, {\sl Phys. Status Solidi} B, {\bf210}, 341.

\item 
{\sc Munteanu, F. M.}, {\sc Kim, Y.}, {\sc Perry, C. H.}, 
{\sc Rickel, D. G.}, {\sc Simmons, J. A.}, and {\sc Reno J. L.}, 
2000, {\sl Phys. Rev.} B, {\bf61}, 4731.

\item 
{\sc Palacios, J. J.}, {\sc Yoshioka, D.}, and {\sc MacDonald, A. H.}, 
1996, {\sl Phys. Rev.} B, {\bf54}, 2296.

\item 
{\sc Quinn, J. J.}, and {\sc W\'ojs, A.}, 
2000, {\sl J. Phys. Cond. Mat.} {\bf12}, R265.

\item 
{\sc Rashba, E. I.}, and {\sc Portnoi, M. E.}, 
1993, {\sl Phys. Rev. Lett.}, {\bf70}, 3315.

\item 
{\sc Shields, A. J.}, {\sc Pepper, M.}, {\sc Simmons, M. Y.}, 
and {\sc Ritchie, D. A.}, 
1995, {\sl Phys. Rev.} B, {\bf52}, 7841.

\item 
{\sc Sitko, P.}, {\sc Yi, S. N.}, {\sc Yi, K. S.}, and 
{\sc Quinn, J. J.}, 
1996, {\sl Phys. Rev. Lett.}, {\bf76}, 3396.

\item 
{\sc Takeyama, S.}, {\sc Kunimatsu, H.}, {\sc Uchida, K.}, 
{\sc Miura, N.}, {\sc Karczewski, G.}, {\sc Jaroszynski, J.}, 
{\sc Wojtowicz, T.}, and {\sc Kossut, J.}, 
1998, {\sl Physica} B, {\bf246}-{\bf247}, 200.

\item 
{\sc Tischler, J. G.}, {\sc Weinstein, B. A.}, and 
{\sc McCombe, B. D.}, 
1999, {\sl Phys. Status Solidi} B, {\bf215}, 263.

\item 
{\sc Turberfield, A. J.}, {\sc Haynes, S. R.}, {\sc Wright, P. A.}, 
{\sc Ford, R. A.}, {\sc Clark, R. G.}, {\sc Ryan, J. F.}, 
{\sc Harris, J. J.}, and {\sc Foxon, C. T.}, 
1990, {\sl Phys. Rev. Lett.}, {\bf65}, 637.

\item 
{\sc Wang, B. S.}, {\sc Birman, J. L.}, and {\sc Su, Z. B.}, 
1992, {\sl Phys. Rev. Lett.}, {\bf68}, 1605.

\item 
{\sc Whittaker, D. M.}, and {\sc Shields, A. J.}, 
1997, {\sl Phys. Rev.} B, {\bf56}, 15~185.

\item 
{\sc Wojtowicz, T.}, {\sc Kutrowski, M.}, {\sc Karczewski, G.}, 
{\sc Kossut, J.}, {\sc Teran, F. J.}, and {\sc Potemski, M.}, 
1999, {\sl Phys. Rev.} B, {\bf59}, R10~437.

\item 
{\sc W\'ojs, A.}, and {\sc Hawrylak, P.}, 
1995, {\sl Phys. Rev.} B, {\bf51}, 10~880.

\item 
{\sc W\'ojs, A.}, {\sc Hawrylak, P.}, and {\sc Quinn, J. J.},
1998, {\sl Physica} B, {\bf256}-{\bf258}, 490;
1999a, {\sl Phys. Rev.} B, {\bf60}, 11~661.
   
\item 
{\sc W\'ojs, A.}, and {\sc Quinn, J. J.}, 
1998, {\sl Solid St. Commun.}, {\bf108}, 493;
1999a, {\sl Solid St. Commun.}, {\bf110}, 45;
1999b, {\sl Acta Phys. Pol.} A, {\bf96} 403;
2000a, {\sl Phil. Mag.} B, {\bf80}, 1405;
2000b, {\sl Phys. Rev.} B, {\bf61}, 2846;
2001a, {\sl Phys. Rev.} B (in press), e-print cond-mat/0006505;
2001b, {\sl Phys. Rev.} B (in press), e-print cond-mat/0007216.
   
\item 
{\sc W\'ojs, A.}, {\sc Quinn, J. J.}, and {\sc Hawrylak, P.}, 
2000a, {\sl Phys. Rev.} B, {\bf62}, 4630;
2000b, {\sl Physica} E, {\bf8}, 254.

\item 
{\sc W\'ojs, A.}, {\sc Szlufarska, I.}, {\sc Yi, K. S.}, and 
{\sc Quinn, J. J.}, 
1999b, {\sl Phys. Rev.} B, {\bf60}, R11~273. 

\end{list}

\end{document}